# CALIBRATION OF CMS CALORIMETERS WITH LHC PROTON BEAM DEFLECTED BY CRYSTAL

Yu.Chesnokov, V.Biryukov, V.Kryshkin, IHEP, Protvino, 142281, Russia

*Abstract*

Calibration of the forward CMS hadron calorimeters *in situ* by the LHC beam is proposed. Simulations show that bent crystal channeling technique is feasible at the LHC, and report the experience of IHEP Protvino in bending 70 GeV protons by 9 degrees (150 mrad) during 10 years in 1994-2004 experiments. Practical realization of calibration scheme based on simulations and previous experience is proposed.

## INTRODUCTION

The goal of the absolute energy scale (about 1%) required for the new generation of the collider experiments is a very tough task. Calibration of hadron calorimeters with fixed target beam is practically impossible:

- in case the calorimeters are supposed to be placed in a magnetic field any calibration outside of magnetic field is a rough approximation because the scintillator and a sandwich calorimeter response depend on magnetic field value and its orientation;

- transportation of calibration coefficients obtained with external beams is not an easy task (requires some corrections connected, for example, with influence of magnetic field on calorimeter response).

Absolute calibration in situ, utilizing some physical processes, is time consuming procedure providing only limited precision (particularly for forward direction due to large number of mini-bias events):

- "in the forward direction Z +*jet* utility becomes more questionable" [1];

- same arguments are applicable to "$\gamma$+*jet*";

- for the process W→*jet* in $t$ - $\bar{t}$ events there is no estimation for forward calorimeters, the pile up can be a serious problem.

Besides a single hadron response of the forward calorimeters is important also if B physics is planned to study. There is no reliable way to do it in situ.

An ideal solution would be to irradiate the assembled calorimeters by high energy beam. There is such beam in hadron collides – circulating proton beam. And the method to direct the beam into the detectors is well known.

## BEAM BENDING METHOD

To irradiate CMS detectors, one would need to bend the beam circulating in the LHC by a huge angle, order of 1-10 degrees (17-170 mrad). Beam intensity as low as 1-1000 p/s would be sufficient for calibration purposes.

Such a bending is possible indeed, but only by extreme, ~1000-Tesla-strong fields of bent crystal. The technique of particle beam channeling by bent crystals is well established at accelerators [2]. Broad experience is obtained with bent channeling crystals at IHEP, CERN SPS, Tevatron, and RHIC over recent decades [2-15]. For instance, bent crystals are largely used for extraction of 70-GeV protons at IHEP (Protvino) [11-15] with efficiency reaching 85% at intensity of $10^{12}$ particle per second, steered by silicon crystal of just 2 mm in length [12]. Much of the IHEP physics program relies on crystal channeled beams regularly used since 1989. Following the successful experiments on crystal channeling in accelerator rings, there has been a strong interest to apply channeling technique in a TeV-range collider for beam extraction or collimation [2-8,16-18].

## IHEP EXPERIMENT ON CRYSTAL BENDING OF 70 GeV PROTONS BY 150 MRAD

The possibility of beam bending by a channeling crystal through very large angles, order of 100 mrad, was studied at Protvino. In 1994-2004 IHEP has operated a channeling crystal 100 mm long to bend 70 GeV beam a huge angle of 150 mrad (9 degrees!) [19,20]. This was done with the purpose to organize over a short base a non-traditional beam line. A 100 mm long Si (110) crystal deflected $10^6$ proton/s beyond the 2-meter iron-concrete shield over 20 m base, out of about $10^{10}$ proton/s hitting the crystal. The crystal efficiency was in agreement with calculation. The orientation-independent component of the signal (background particles) did not exceed 3% of the channeled beam. The scheme of 150-mrad crystal beam line operated at IHEP at $10^6$ proton/s in 1994-2004 is shown in Figure 1. This crystal beam line was used mostly for our studies of channeling. It consumed practically no power and allowed one to work in parallel with other beam lines without affecting other physical set-ups operation.

Besides this 150-mrad bent crystal used in 1994-2004, another example in IHEP experience was an 85-mrad bent crystal used for extraction in 1989-1999 over 10 years until a new crystal replaced it [2].

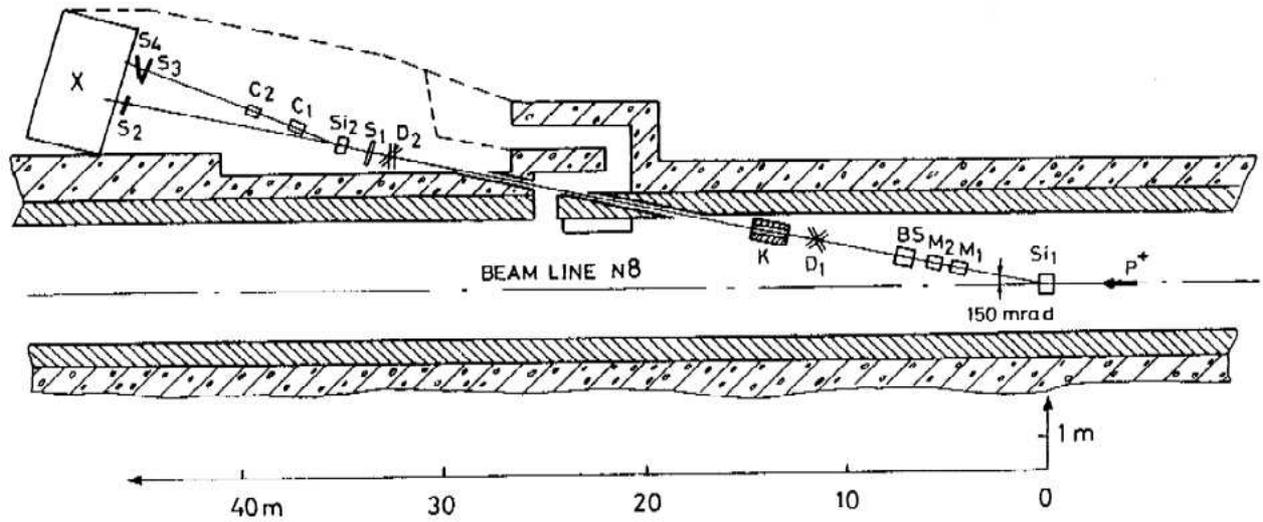

Fig. 1. The scheme of crystal beam line and experimental setup: $Si_1, Si_2$ – deflecting and testing crystals, $M_1, M_2$ – corrector magnets, BS – beam stopperp, $D_1, D_2$ – proportional chambers, K – a collimator, $S_1 - S_4$ – scintillator countes, C1, C2 – microstrip detector stations, X – a beam absorber.

## SIMULATIONS FOR THE LHC BEAM BENDING BY 1-20 DEGREES

In order to find out whether one can extrapolate the 70 GeV experience onto the LHC case of much higher energy range, we performed Monte Carlo simulations applying the code [21-23] used in many other channeling experiments [6-15]. Details of the theoretical calculations can be found, e.g., in book [2].

Our preliminary study assumed a crystal of 50 cm size, which is a reasonable figure available on the market in Russia and Europe. Earlier, IHEP has used up to 15 cm long Si crystals for channeling.

An example of the calculation of efficient transmission of LHC beam with Silicon crystal at the energies of 450 GeV (injection level in the LHC) and 1000 GeV is shown in Figure 2. Respectively, bending angles up to about 7 and 11 degrees can be realized with Si crystal at these energies.

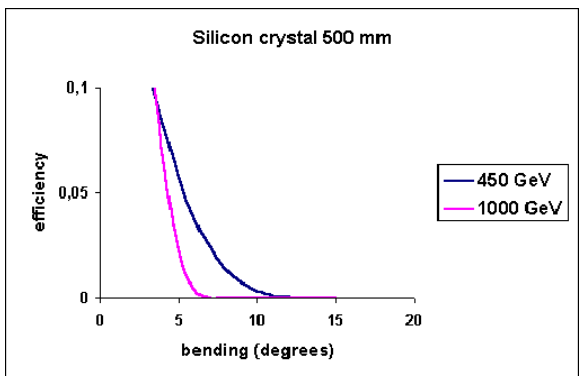

Figure 2 Efficiency of the LHC proton beam bending with a single *Si* crystal 500 mm long as a function of the bending angle (degrees).

We have run computations for Silicon and Germanium crystals of 50 cm size and the LHC energy range from injection, 450 GeV, through the top energy, 7 TeV. Some preliminary results are shown on Figure 3 for the bending angle achievable with a single Si crystal of 50 cm length as a function of the LHC beam energy.

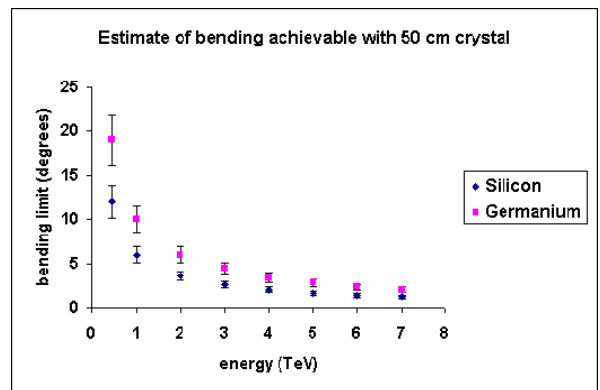

Figure 3 Bending angle achievable with a single Si crystal of 50 cm length as a function of the LHC beam energy.

Thus simulations show: if the beam energy is limited to the injection energy (450 GeV) that is sufficient for reliable measurement of the energy scale the size of the crystal can be a modest one.

All towers of the hadron calorimeters (HE and HF) are irradiated by a radioactive source and relative coefficients are known for all towers therefore in principle it is sufficient to define an energy scale only for a single tower. But due to harsh radiation environment in forward direction the active calorimeter elements can be damaged. Therefore control of the tower performance is required in all available psevdorapidity range (no scanning in $\varphi$ is required because of the axial symmetry). The solution of the task is depicted in fig. 4. Not all channeled particles are bent to a maximum angle. Positive feature of large crystal bending for calibration is wide angular region of the deflected particles: there are present not only fully deflected particles, but also some particles dechanneled in a crystal bulk and thus deflected at smaller angle.

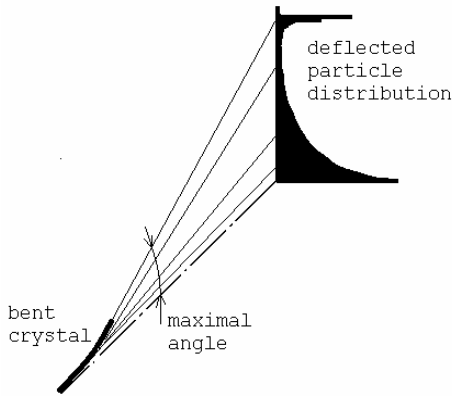

Fig. 4. Angular distribution of the bended particles.

## LAYOUT OF A CALIBRATION SCHEME

Example of the realization of the proposed method is illustrated in fig. 5 where the central part of CMS is shown. A crystal can be placed inside of the electromagnetic calorimeter EE. Then the maximum particle bending angle will be about $12^0$ and the proton beam will strike HE tower far enough from the inner edge to avoid leakage into HF. On the other hand the combined performance of HE and HF can not be studied with extracted beams and the use of the channeled crystal inserted in beam is the only way to measure this characteristic.

The place inside of the EE is shown in fig. 6. There is a bellows to join two parts of the beam pipe and no special efforts are needed to insert a crystal. The maximal available *Si* crystal length is about 350 mm. According to the fig.6 after removing the part of the pipe with the bells there is a free space 410 mm length.

The bent crystal can be fixed in a Π shape aluminum profile presented in fig.7 and positioned in the beam pipe (fig. 8). There exist several possible options to direct the particles at the crystal which depend on the dynamics of the beam. For example if the injected 450 GeV beam has appreciable halo the crystal can be placed at such distance from the axis that the beam halo will overlap the crystal. This period can be used for calorimeter calibration. After the proton accumulation the beam halo is decreased and the crystal will be outside of the beam. The other way is to use existing magnetic elements to steer the beam into the crystal. No moving parts (goniometer and so on) are needed due to application of special crystal with triangular shape of entry face (fig. 9). Such crystal type has distributed acceptance and its initial alignment with an accuracy of about few milliradian is enough for channeling achievement. At fig. 10 the diminish-scale model of necessary crystal device is presented. The full scale device can be constructed and tested in 2006 test beam time with HE and HF prototypes.

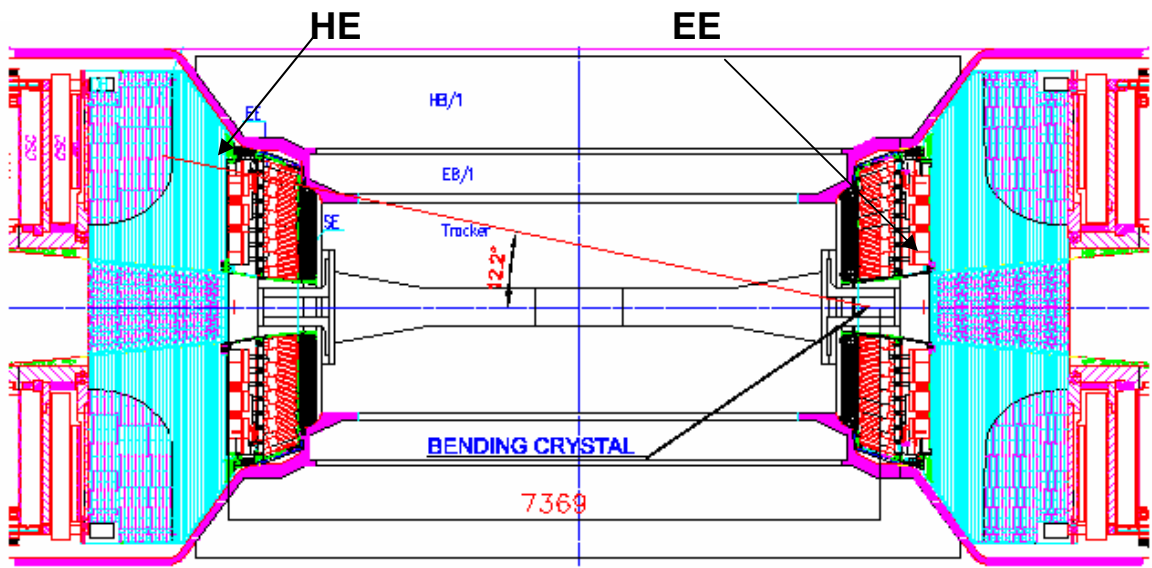

Fig. 5. Layout of the central part of CMS with crystal placed inside electromagnetic calorimeter EE.

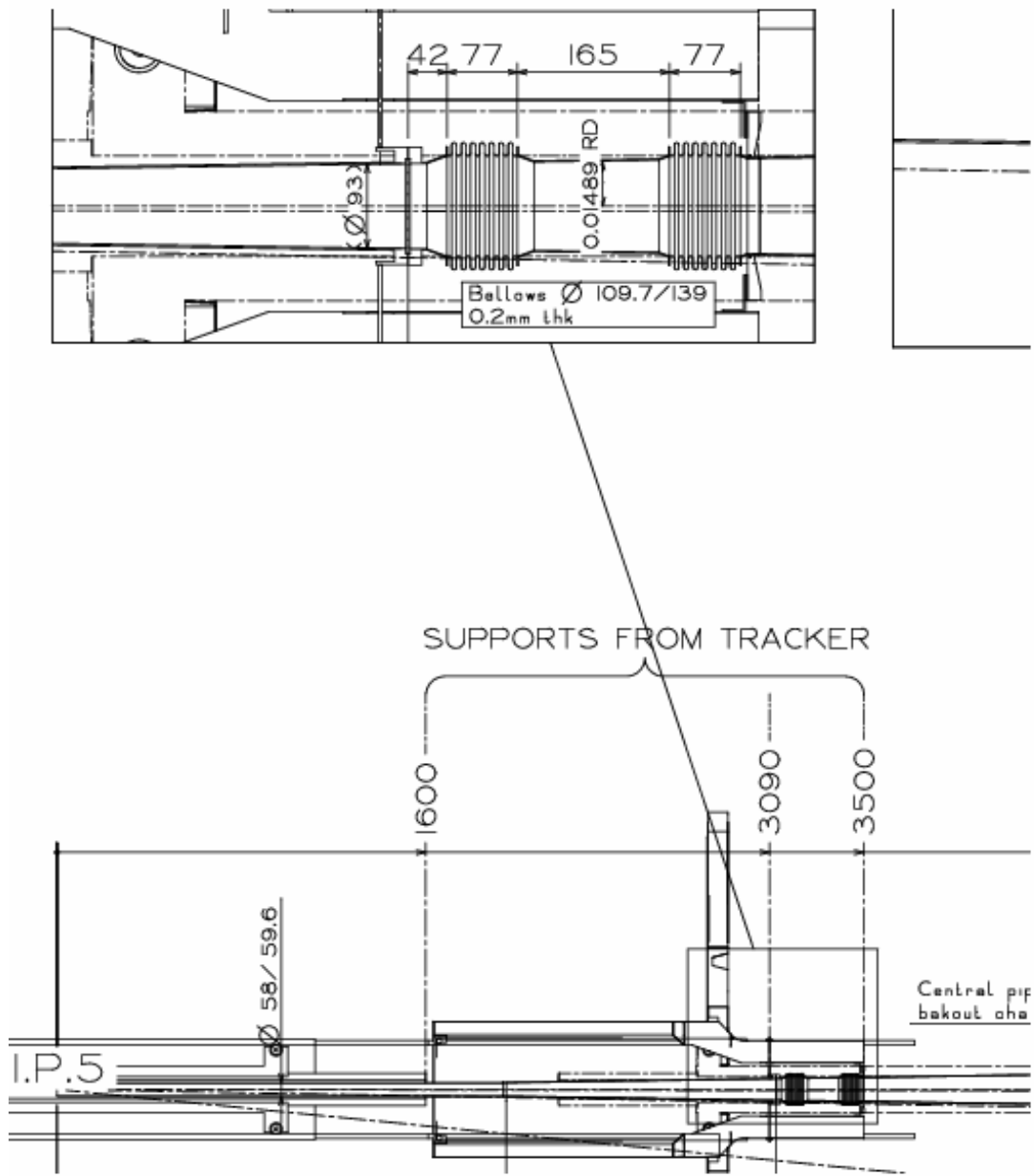

Fig. 6. Layout of the beam pipe inside of the EE. I.P.5 is the interaction point.

Fig. 7. Crystal holder made of aluminum Π shape profile.

Fig. 8. Example of a crystal holder installation inside of the beam pipe.

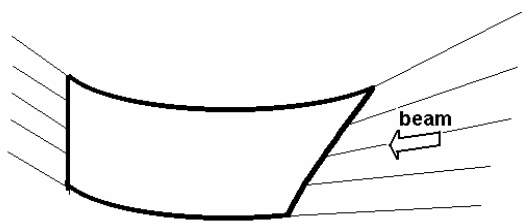

Figure 9. Scheme of crystal with skew entry face. The tangents to crystalline planes at crystal entry are non parallel in the limits of few milliradian witch excluded the necessary of crystal rotation for achievement of channeling mode.

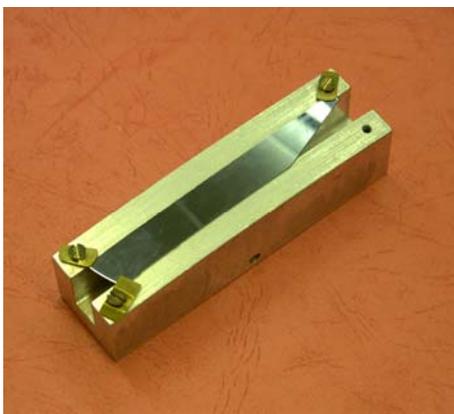

Figure 10. The diminish scale model of bent crystal with skew entry. Model parameters are: 150 mrad bent, 100 mm length and 12 mm width.

## CONCLUSION

The simple method can be used for absolute calibration in real environment. Unique characteristic (such as the combine performance of HE and HF) can be thoroughly studied. The calibration can be done regularly (at least each day). So it can be used for the control also (variation of scale and resolution with time). No radioactive sours can give this information (if some channel failed the resulting degradation of the energy resolution can not be calculated).

For HE calibration the maximum required angle is about $12^0$ that is close for the present achievements. As for HF the bending angle is well within the currently used range and it will not require additional study.